\documentclass[9pt,twocolumn,twoside]{optica}
\setboolean{shortarticle}{true}
\setboolean{minireview}{false}
\usepackage{graphicx,amssymb,mathtools,bm,nicefrac,blindtext,mathptmx} 

\newcommand{\openone}{\leavevmode\hbox{\small1\normalsize\kern-.33em1}} 

\newcommand{\bra}[1]{\left\langle #1 \right|}
\newcommand{\ket}[1]{\left| #1 \right\rangle}
\newcommand{\braket}[2]{\left\langle {#1{\left| \vphantom{#1 #2} \right.} #2} \right\rangle}

\renewcommand{\epsilon}{\varepsilon}

\newcommand{\Tr}{\mathop{\mathrm{Tr}} \nolimits}
\newcommand{\lmax}{{\ell_{\mathrm{max}}}}

\title{Compressed sensing of twisted photons}

\author[1]{F.~Bouchard} 
\author[2]{D.~Koutn{\'y}}
\author[1]{F.~Hufnagel}
\author[2]{Z.~Hradil}
\author[2]{J.~\v{R}eh\'a\v{c}ek}
\author[3]{Y.~S.~Teo}
\author[3]{D.~Ahn}
\author[3]{H.~Jeong} 
\author[4,5,*]{L. L.~S\'{a}nchez-Soto} 
\author[1,5]{G.~Leuchs} 
\author[1,5]{E.~Karimi}

\affil[1] {Department of Physics, University of Ottawa, 25 Templeton Street, Ottawa, Ontario, K1N 6N5 Canada}

\affil[2]{Department of Optics, Palack\'y University, 17. listopadu 12, 771 46 Olomouc, Czech Republic}

\affil[3]{Department of Physics and Astronomy, Seoul National University, 08826 Seoul, South Korea}

\affil[4]{Departamento de \'Optica, Facultad de F\'{\i}sica, Universidad Complutense, 28040~Madrid, Spain} 

\affil[5]{Max-Planck-Institut f\"ur die Physik des Lichts, Staudtstra\ss e 2, 91058 Erlangen, Germany}

\affil[*]{Corresponding author: lsanchez@fis.ucm.es}

\dates{Compiled \today}

\ociscodes{(270.0270) Quantum optics; (270.5585) Quantum information and processing; (120.3940) Metrology.}

\doi{\url{http://dx.doi.org/10.1364/optica.XX.XXXXXX}}

\begin{abstract}
  The ability to completely characterize the state of a quantum system
  is an essential element for the emerging quantum technologies. Here,
  we present a compressed-sensing inspired method to ascertain any
  rank-deficient qudit state, which we experimentally encode in
  photonic orbital angular momentum. We efficiently reconstruct these
  qudit states from a few scans with an intensified CCD camera. Since
  it requires only a few intensity measurements, our technique would
  provide an easy and accurate way to identify quantum sources,
  channels, and systems.
\end{abstract}

\setboolean{displaycopyright}{true}

\begin{document}

\maketitle
\thispagestyle{fancy}

\ifthenelse{\boolean{shortarticle}}{\abscontent}{}

The orbital angular momentum (OAM) of single photons, which provides
an unbounded vector space, has been recognized as a preeminent
platform for encoding both
quantum~\cite{Vallone:2014aa,Mirhosseini:2015aa,Sit:2017aa} and
classical~\cite{Gibson:04,Wang:2012aa} information. Although the
generation of photons with OAM is relatively simple, the full
characterization of a quantum state in the OAM Hilbert space stands as
a challenging task. Several methods have demonstrated accurate
projective measurements to determine OAM
states~\cite{Mair:2001aa,Leach:2002aa,Leach:2004aa,Karimi:2009aa,
  Berkhout:2010aa,Mirhosseini:2013aa}. However,
these projective measurements work adequately only for pure
states. The case of mixed states requires full state tomography, and
this involves projective measurements on arbitrary superpositions of
two or more OAM eigenstates~\cite{Bent:2015aa}, a task which remains
challenging.

To circumvent these problems, we propose and
experimentally demonstrate a method inspired by compressed
sensing. Originating from the context of classical signal
processing~\cite{Candes:2006aa,Candes:2010aa},
this technique harnesses prior assumptions about the state to
reconstruct it from an undersampled set of measurements. It is
routinely used to estimate vectors or matrices from incomplete
information, with applications in many diverse fields of
research~\cite{Eldar:2012aa,Stern:2016aa}. Compressed
sensing has also been adapted as a tool for state tomography of
discrete systems in quantum theory~\cite{Gross:2010aa,
  Shabani:2011aa,Liu:2012aa,Schwemmer:2014aa,Rodionov:2014aa,
  Tonolini:2014aa,Steffens:2017aa,Riofrio:2017aa}. Our scheme involves
making a small number of intensity scans with an intensified CCD
camera. The compressed sensing algorithms, supplemented with the
positivity constraint~\cite{Kalev:2015aa}, which operates as a kind of
regularization, enables us to construct informationally complete
measurements that are robust to noise and modeling errors. In other
words, one can use a very simple setup for the characterization of a
quantum state with \emph{a priori} information about its nature.

We begin our analysis by recalling a few basic concepts that we will
need to understand the technique. In any tomographic protocol, one
infers the quantum state, represented by the density matrix $\varrho$,
from the distinct outcomes of a collection of measurements performed
on identical copies of the system.  The outcomes of these measurements
are given by the Born rule $p_{\alpha}=\Tr(\varrho\,\Pi_{\alpha})$,
where $\{\Pi_{\alpha}\}$ is the positive operator-valued measure
(POVM) describing the setup~\cite{Peres:2002oz}.  We denote the action
of the POVM by $\mathcal{A}:\varrho \mapsto \mathbf{p}$, that maps
$\varrho$ onto the vector containing all the probabilities
$\{p_{\alpha} \}$.

\begin{figure*}[t]
  \centerline{\includegraphics[width=1.95\columnwidth]{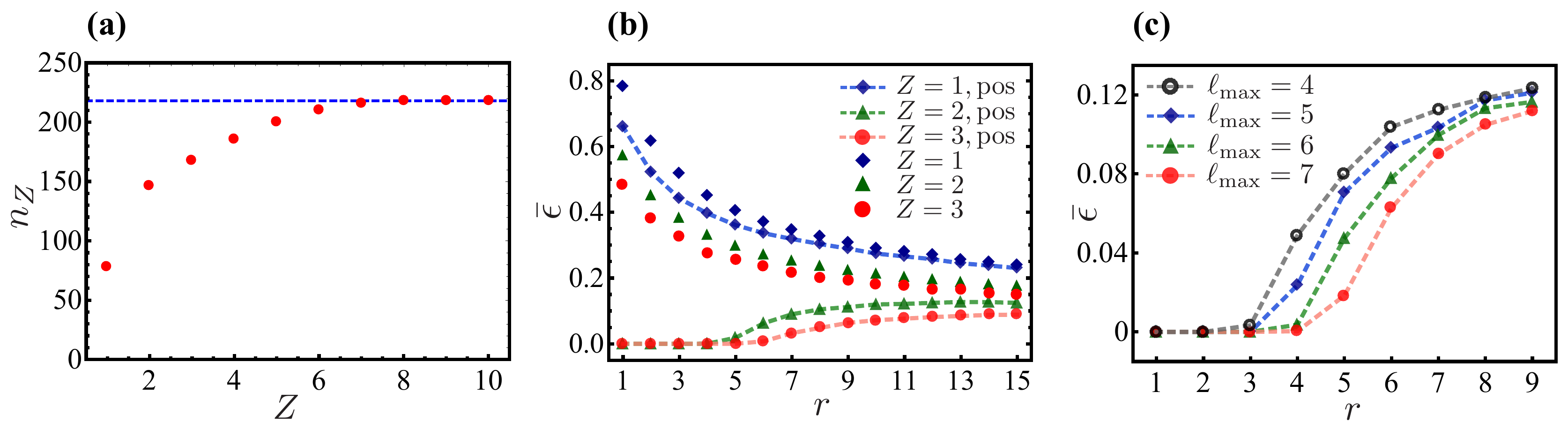}}
  \caption{Simulation of our compressed sensing protocol with twisted
    photons. (a) Number of independent detections $n_Z$ generated by
    $Z$ intensity scans performed on a signal with $\lmax=7$
    ($d=15$). The maximum of $n_Z=218$ detections is obtained for
    $Z \ge 8$ scans. (b) Reconstructions errors, $\bar{\varepsilon}$,
    from the compressed sensing protocol of the twisted photons as a
    function of the rank of the state for a fixed dimension
    ($\lmax = 7$) and for $Z=1, 2,$ and 3 CCD scans, with and without
    positivity constraint. (c) The reconstructions errors for two CCD
    scans and different dimensions. In all the cases, we take a
    $19\times19$ pixels screen.}
  \label{fig:zup}
\end{figure*}

A POVM is informationally complete (IC) when the corresponding outcome probabilities are sufficient to determine an arbitrary
state~\cite{Prugovecki:1977fk,Busch:1989kx}. Obviously, the number of outcomes of any IC measurement is at least $d^{2}$, which makes traditional methods infeasible as the dimensionality of the system increases. However, if we know \emph{a priori} that  the rank of the system fulfills $\mathrm{rank}(\varrho)\leq r$, with $r\ll d$, then we can substantially reduce the number of  measurement samples required to uniquely reconstruct the unknown signal matrix. This can be accomplished with a map $\mathcal{A}$ that satisfies the appropriate properties~\cite{Candes:2008ab}; the true state $\varrho_{0}$ is then the only density matrix within the set of positive Hermitian matrices of any rank that yields the measurement probabilities $\mathbf{p}$~\cite{Kalev:2015aa}. The corresponding estimator $\widehat{\varrho}$ is given by,
\begin{equation}
  \label{eq:constopt}
	\begin{split}
          \widehat{\varrho} =  \min_{\varrho} \|
          \mathcal{A} [\varrho]-\mathbf{p} \|
          \quad \mathrm{s. \, t.} \quad \varrho \geq 0 \, ,
	\end{split}
\end{equation}
whose solution can be efficiently found by convex programming~\cite{Boyd:2004qd}.

The insights into the regularizing effect of the positivity constraint permit us to tackle reconstructing  OAM quantum states. To represent the structure of the transverse field we will be using the well-known Laguerre-Gauss (LG) modes that can be written as~\cite{Siegman:1986aa},
\begin{eqnarray}
  \label{eq:LG}
  \mathrm{LG}_{p \ell}(r, \phi,z) =
  \braket{r, \phi, z}{\ell, p} = \sqrt{\frac{2p!}{\pi (p + |\ell|)!}}
  \frac{1}{w(z)} \left (\frac{\sqrt{2} r}{w(z)} \right )^{|\ell|}  \nonumber \\
  \times L^{|\ell|}_{p} \left(\frac{2r^{2}}{w(z)^{2}}\right)
  \exp{\left( -r^{2}  \left [ \frac{1}{w(z)^2} -
  i \frac{k}{2R(z)} \right ] -i \ell \phi - i \psi_{p \ell} (z) \right)},
\end{eqnarray}
where $(r,\phi,z)$ denote cylindrical coordinates, $k$ is the wave
number, $L_{p}^{|\ell|} (.)$ is the generalized Laguerre polynomial,
$\ell \in \{ 0, \pm 1, \pm 2, \ldots \}$ is the azimuthal mode index,
and $p \in \{ 0, 1, 2, \ldots \}$ is the radial mode index, which is
related to the number of radial nodes. The parameters $R(z)$, $w(z)$,
and $\Psi_{p\ell} (z)$ denote the radius curvature of the wave fronts,
the beam radius, and the Gouy phase at the propagation distance $z$,
respectively:
$w^{2}(z) = w^{2}_{0} \left [ 1 + ( z/z_{\mathrm{R}})^{2}\right ] $,
$R(z) = z\left[ 1 +(z_{\mathrm{R}}/z)^{2}\right ]$, and
$\psi_{p\ell} (z) = (2 p + |\ell|+ 1) \arctan (z/z_{R})$, with the
Rayleigh range $z_{\mathrm{R}} = kw_{0}^{2}/2$ and $w_{0}$ the beam
waist, which we assume located at $z=0$. In what follows, we will set
$p=0$ and denote $\psi_{\ell} (z) = \psi_{p=0 \, \ell} (z)$.

As mentioned before, we propose to reconstruct the signal in the LG
basis from $Z$ intensity scans registered by a CCD camera positioned
at distances $z_1, z_2, \ldots, $ with respect to the beam waist,
which is in the spirit of the Gerchberg-Saxton
algorithm~\cite{Gerchberg:1972aa}. This amounts to projecting the
density matrix on free-space position eigenstates; viz,
\begin{equation}
  \label{eq:comb}
  p(r,\phi,z)=\langle r,\phi,z| \varrho| r,\phi,z \rangle \propto
  e^{-2\mathfrak{r}^{2}} \sum_{\ell \ell^{\prime}}
  \varrho_{\ell\ell^{\prime}} \,
  C_{\ell \ell^{\prime}},
\end{equation}
where $C_{\ell \ell^{\prime}} = \mathfrak{r}^{|\ell|+|\ell^{\prime}|}
\exp[i(\ell-\ell^{\prime})\phi] \exp \{i[\psi_{\ell}
(z)-\psi_{\ell^{\prime}} (z)]\} $ and $\mathfrak{r} (z) = r/w(z)$.  Each
combination of $r$, $\phi$, and $z$ coordinates, which corresponds to
the associated pixel readings, gives us one particular linear
combination of density matrix elements. Evidently by varying $r,\phi$,
and $z$, the more linearly independent these generated combinations
are, the more IC the POVM is.

Let us first consider $d$-dimensional states that can be represented
in the LG mode basis with nonnegative azimuthal indices only; i.e.,
$\ell\in\{0,1,\ldots,d-1\}$. Incompleteness arises whenever two
different pairs $(\ell^{\prime},\ell^{\prime \prime})$ generate the
same $C_{\ell^{\prime}\ell^{\prime \prime}}$ for all $r$ and
$\phi$. Surprisingly, it turns out (see Supplemental Material) that
all density matrix elements can be uniquely determined from a single
CCD image for any dimension $d$. Put differently, a simple von Neumann
measurement (CCD scan) defined in an infinitely large space, becomes
an IC POVM when projected into a finite space.

Next, we examine the case in which the sign of the topological charge
is not constrained; that is, $d=2 \lmax+1$, with
$\ell\in\{-\lmax,\cdots,0,\cdots,\lmax\}$. CCD scans are no longer IC,
no matter how many planes $Z$ are sampled. As discussed in the
Supplemental Material, the maximum number of linearly independent
measurements generated by CCD scans is
$n_{Z \rightarrow\infty}=d^2- (d-1)/2$. Only for large dimensions,
$d^2\gg1$, does the measurement become nearly IC. The first two scans
generate $ n_{Z=1} = (d^2+6d-3)/4$ and $ n_{Z=2} = (d^2+5d-8)/2$
independent detections. For higher $Z$, the new independent detections
grow linearly with $\lmax$. A typical behaviour is shown in
Fig.~\ref{fig:zup}a) for $\lmax=7$: the maximum
$n_{Z \rightarrow\infty}$ is approximately attained with $8$
scans. Actually, one can determine (see Supplemental Material) that
the minimum scans required to approximately attain the maximum number
of independent detections in the signal space is
$Z_\mathrm{min}=\lmax+1$. Roughly one quarter and one half of this
total is already available after one and two scans, respectively.

The compressive tomography is done by solving \eqref{eq:constopt},
where now $\alpha=\{r,\phi,z\}$, and $\Pi_{\alpha}$ is defined by
$\Tr ( \rho\,\Pi_{\alpha})= p(r,\phi,z)$. We first perform numerical
simulations to explore the method. A set of random states of a given
rank in the $\lmax=7$ space is chosen and scanning measurements are
simulated for $Z=1$, $2$, and $3$ CCD planes with a $19\times19$ pixel
geometry.

\begin{figure}[h]
  \centerline{\includegraphics[width=0.87\columnwidth]{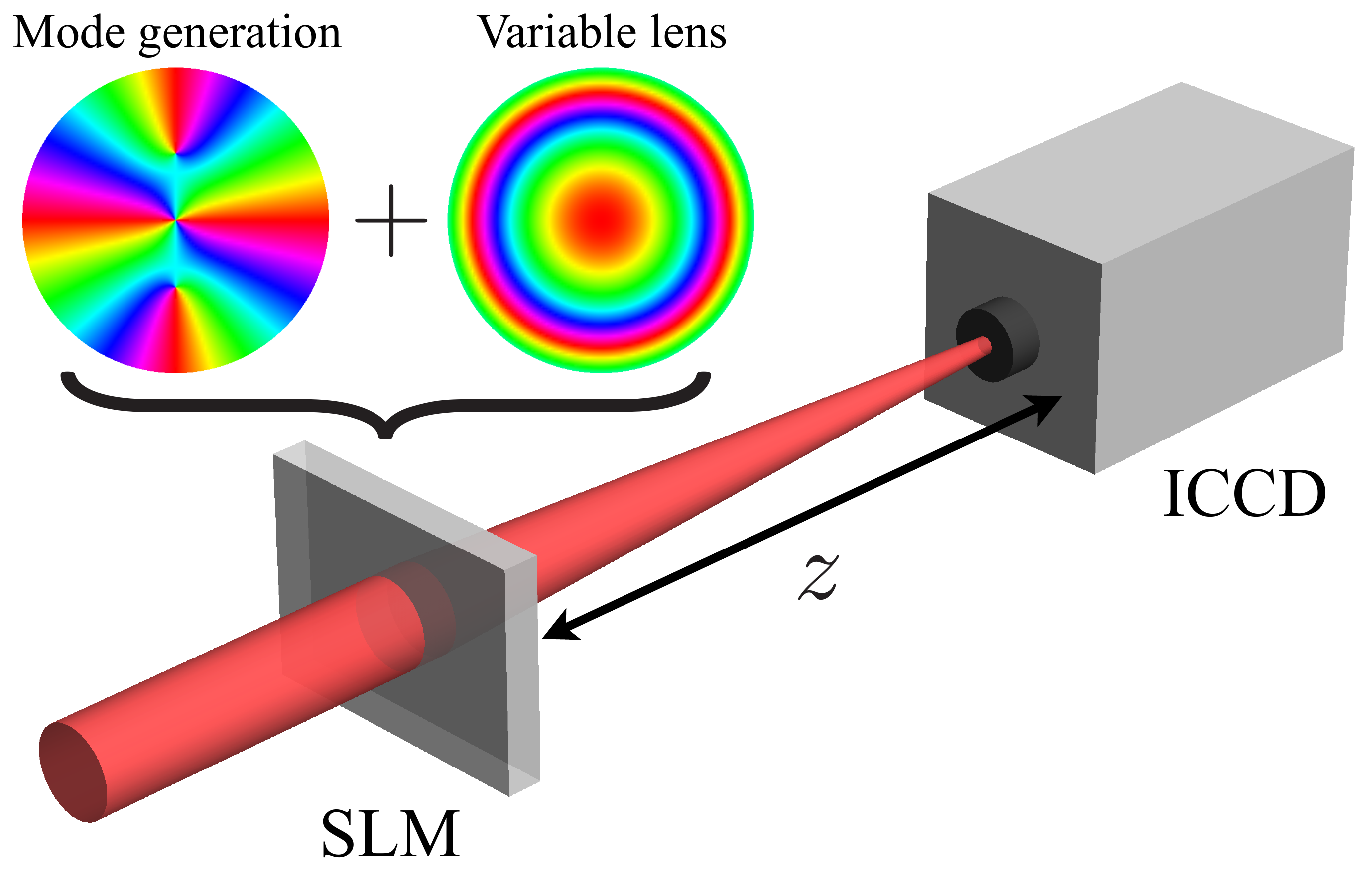}}
  \caption{Sketch of the experimental setup. A photonic state is
    generated by manipulating the phase and intensity of an incoming
    beam via the spatial light modulator (SLM). The beam is then
    focused using a variable holographic lens imprinted on the SLM
    together with the state generation hologram. The ICCD camera has a
    fixed position and records intensity scans.  Inset shows the state
    and lens phase patterns, $[0,2\pi)$, in a hue color.}
  \label{fig:setup}
\end{figure}

\begin{figure*}[t]
  \centerline{\includegraphics[width=1.90\columnwidth]{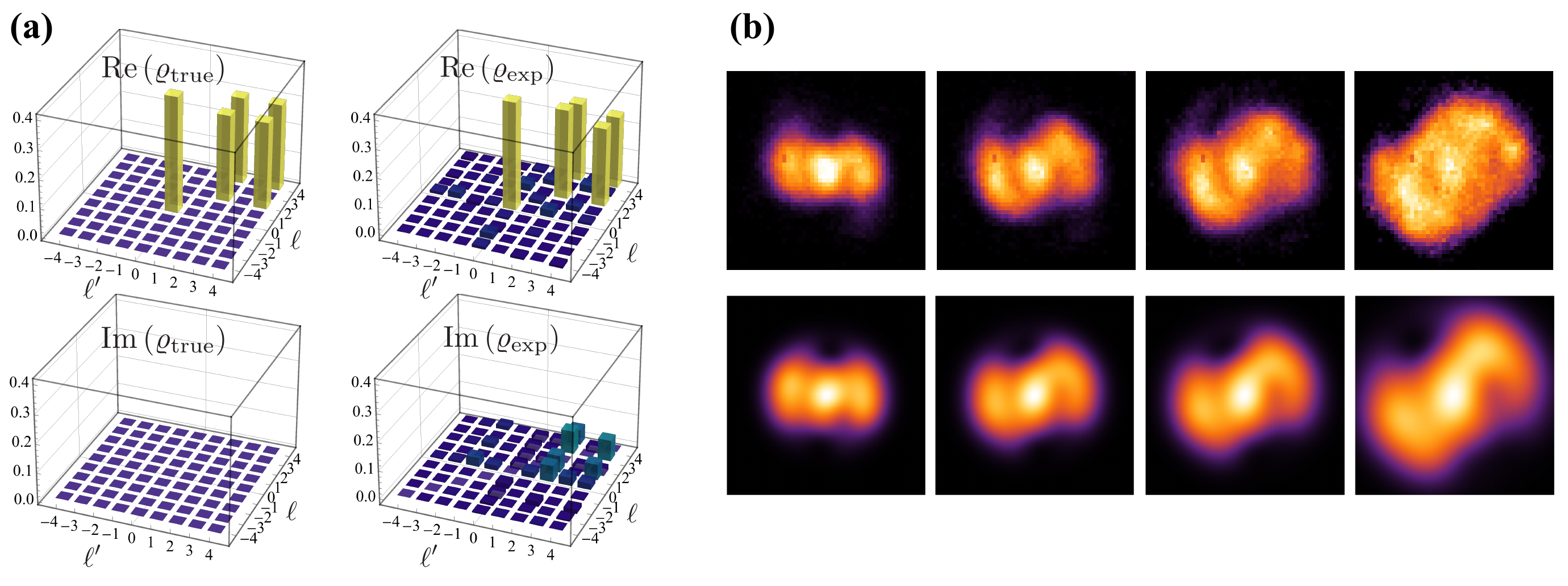}}
  \caption{Experimental reconstruction from compressed sensing. (a)
    Density matrix of the true state and (b) Reconstructed density
    matrix after two intensity scans with a signal space spanned by
    $\ell \in \{-4, \dots, 4\}$.  The upper row shows four
    experimental ICCD scans at the planes $z/z_{\mathrm{R}}=$0, 1/3,
    1/2, and 1, respectively. The lower row shows the predictions from
    the reconstructed state of the same ICCD scans at before.}
  \label{fig:rho1_pos}
\end{figure*}
The data is subject to quantum state reconstruction with and without
the positivity constraints, where the latter is implemented with the
help of the Moore-Penrose pseudoinverse~\cite{Ben-Israel:1977aa}.
Finally, for each true state, the corresponding Hilbert-Schmidt
distances between the true and reconstructed states are calculated and
the mean is taken over the set of true states of the same rank. We
take this as the error $\bar{\varepsilon}=\mathbb{E}[ \Tr ( \varrho_{\mathrm{true}}-\widehat{\varrho})^2]$.

In Fig.~\ref{fig:zup}b), we have plotted the errors
$\bar{\varepsilon}$ as a function of the rank of the state, for
$\lmax=7$.  We observe a strong regularizing effect of the positivity
constraint, especially on low-rank signals. First, positivity always
makes the reconstruction errors smaller. Second, it cures the
informational incompleteness of the tomography, provided the
complexity of the measured signal stays below a certain
threshold. Last, this threshold gets increased as we collect more
scans. Notice though that protocols with positivity perform worse on
more complex signals because the regularizing effect of positivity
weakens. Loosely speaking, in comparing neighbourhoods of low- and
high-rank states, one finds more nonphysical matrices in the
former. Those objects get filtered by the positivity constraints
leading to improved performance. The opposite is true of the protocols
without positivity. Here, the more mixed states are less biased in the
space of density matrices, so their average distance to the
reconstructed matrix is smaller. Notice that this discussion does not
apply to mutually compatible observations, for which positivity
constraints are of no consequence. This can be observed in
Fig.~\ref{fig:zup}c) for a single scan, where the regularizing effect
of positivity disappears. Stronger incompatibilities introduced by
detections from different planes
$[\Pi_{\alpha}, \Pi_{\alpha^{\prime}}] \neq 0$ are required to promote
signal sparsity. The influence of the signal dimension with a fixed
measurement is also illustrated in Fig.~\ref{fig:zup}c), for the
particular case of two scans. The quality of the compressive OAM
tomography improves with the signal dimension. At first glance, this
might seem counterintuitive; however, note that the sparsity of a
signal of a given rank grows with dimension and so do the regularizing
effects of positivity constraints.

To verify our scheme, we construct an experimental setup that allows
us to generate and detect vortex beams at the single photon level
using a spatial light modulator (SLM) and an intensified CCD (ICCD)
camera, respectively. A simplified sketch is shown in
Fig.~\ref{fig:setup}. A quasi-continuous UV laser, at a wavelength of
355~nm, is used to pump a type-I $\beta$-barium borate (BBO) crystal
in order to generate single photons via spontaneous parametric
downconversion. The spatial modes of the generated photons are
filtered to the fundamental Gaussian mode by coupling them to a single
mode fibre (SMF). The photons are then coupled out of the SMF and are
made incident on an SLM, (X10468-07, Hamamatsu) consisting of an
electronically controlled nematic liquid crystal device with
$792 \times 600$ pixels. The phase and the intensity of the generated
photons are controlled via a holographic intensity masking
technique~\cite{bolduc:2013aa}, with a diffraction efficiency greater
than 70~\%. To achieve a higher mode quality, a 10$\times$ microscope
objective with a numerical aperture of 0.25 is employed to obtain a
relatively large collimated beam at the SLM. The mode waist of the
amplitude-modulated hologram at the SLM is determined to have an
effectively flat phase and intensity for the incoming beam over the
region of interest of the hologram. By doing so, we may generate any
arbitrary spatial mode with a high-level of accuracy. In the case of
pure states, a single hologram is used for shaping the transverse
modes of all photons in the ensemble. For mixed states, the
appropriate holograms are randomly varied to generate incoherent
statistical mixtures of the desired modes. A single intensity scan is
then recorded over the whole ensemble of generated photons.

To observe the intensity variation of LG modes upon propagation, due
to the differences in Gouy phases, the beam must propagate for
distances that are on the order of the Rayleigh range
$z_{\mathrm{R}}$. This is achieved by simply focusing the beam. In the
scheme described above, the ICCD camera is moved along the path of the
beam in the $z$-direction. However, this is equivalent to varying the
focal length of the lens and considering a fixed position of the ICCD
camera. This is accomplished by imprinting a phase profile of the form
$\exp\left(-i\,k\,r^2/2f\right)$, where $f$ is the focal length of the
flat lens. By doing so, our experimental reconstruction does not
require any mechanical displacement of the components in the
setup. Intensity scans are recorded for different focal lengths
corresponding to positions of $z/z_\mathrm{R}=0$, 1/3, 1/2 and 1.

\begin{figure}[t]
  \centering{\includegraphics[width=0.8\columnwidth]{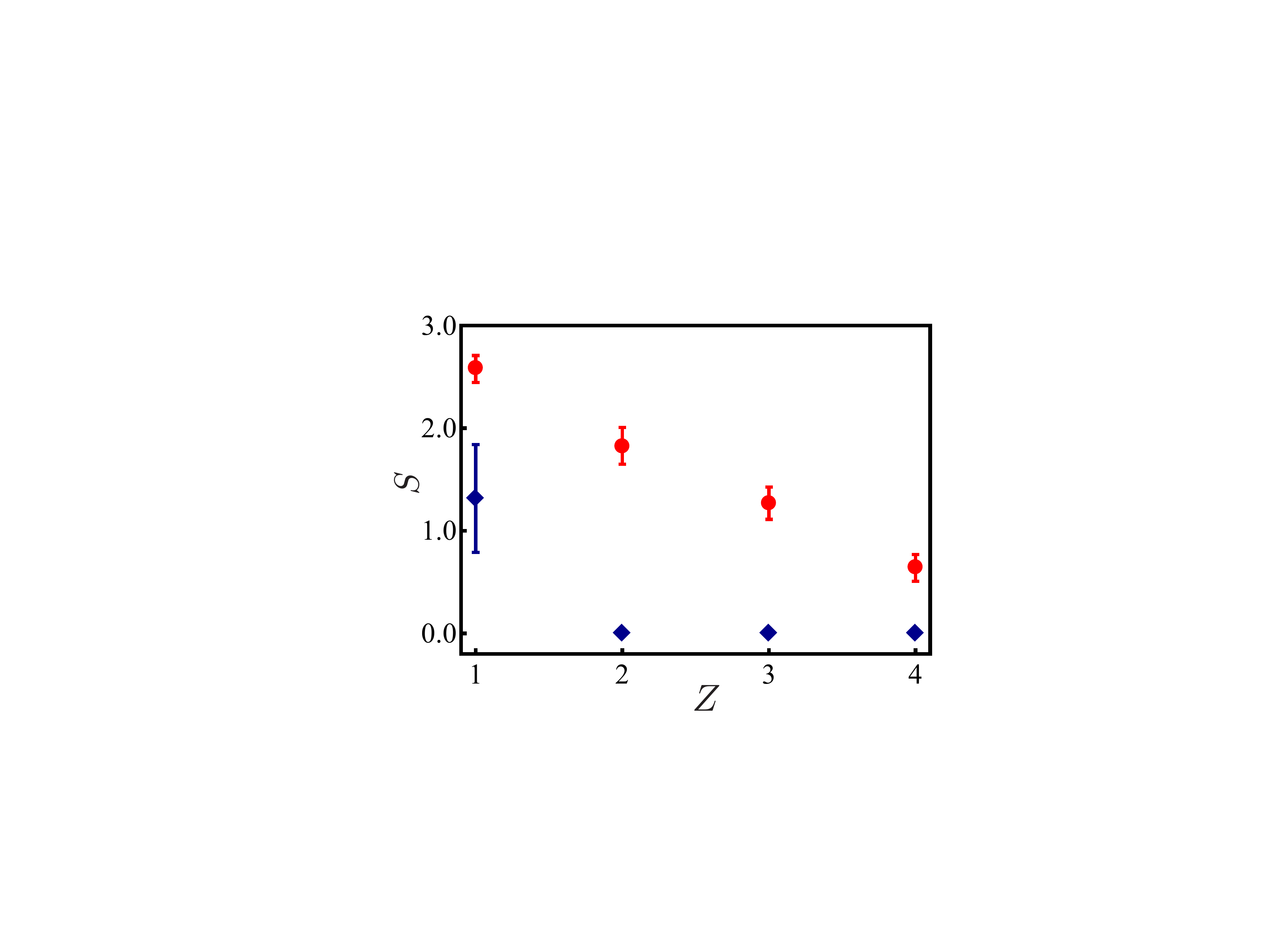}}
  \caption{Entropy $S$ as a function of the number of intensity scans
    $Z$, for a signal space spanned by $\ell \in \{ -4,\dots,4\}$ and
    for true states of the form in \eqref{eq:r1}. The
    circles are obtained by averaging over 20 random states and the
    error bars indicate the corresponding variance. Blue refers to
    protocol with positivity, while red is without positivity.}
  \label{fig:ent}
\end{figure}

The true states are chosen to be following rank--$2$ states,
\begin{eqnarray}
  \label{eq:r1}
  \varrho= p\ket{0}\bra{0}+\left(1-p\right) \ket{\Psi}\bra{\Psi},
\end{eqnarray}
with $\ket{\Psi}=\cos\theta\, \ket{-3} + \sin\theta \, \ket{3}$. In
the experiment, we randomly generate 20 states for values of $p$ and
$\theta$ chosen. The experiment was also carried out with other sets
of states; the results can be found in the Supplemental Material. In
Fig.~\ref{fig:rho1_pos}, we present the results for a typical state
$\varrho$ with two ICCD scans. The corresponding reconstructed density
matrix is plotted in the left column. Even though we are dealing with
informationally incomplete measurements, the protocol with positivity
using two intensity scans fits the theoretical data quite
well. Without the positivity constraint, the reconstructed states are
very different from the true ones. We have also experimentally checked
that when the signal space is constrained to only positive topological
charges $\ell$, one intensity scan is enough to accurately predict the
intensity scans in other planes, as anticipated by our theory. To
assess the degree of IC for the intensity scans, we first consider a
matrix of reconstructed states $\{\widehat{\rho}_j\}$ with its $j$th
column defined as a flattened $\rho_j$. A unique solution for the
extremal problem (\ref{eq:constopt}) then implies that the entropy of
normalized singular values for this matrix is zero. To compute the
entropy, we reconstructed 20 true states for every ICCD scan, and
average over all of them. As can be seen in Fig.~\ref{fig:ent}, the
protocol with positivity constraints is enough to characterize the
input state with solely two scans. We corroborate once again that if
we restrict our space to be spanned only with nonnegative azimuthal
indices, the measurement is IC for both strategies (with and without
positivity) with only one intensity scan.

In summary, we have developed a compressed sensing scheme able to
uniquely reconstruct any rank-deficient qudit state encoded in the OAM
degree of freedom. The positivity constraint has played a substantial
role as a powerful regularization to perform a tomographic
reconstruction in the regime of informationally incomplete data for
intermediately sized quantum systems. This establishes a novel and
efficient tomographic paradigm for OAM systems that could trigger
interesting experimental research on complex quantum states, which
otherwise might have been infeasible with currently known detection
schemes in such experiments.

This work was supported by the European Union's Horizon 2020 Research and Innovation Programme (Q-SORT) grant number 766970, Canada Research Chairs (CRC), the BK21 Plus Program (21A20131111123) funded by the Ministry of Education (MOE, Korea) and National Research Foundation of Korea (NRF), the NRF grant funded by the Korea government (MSIP) (Grant No. 2010-0018295), the Korea Institute of Science and Technology Institutional Program (Project No. 2E27800-18-P043), the Spanish MINECO (Grant FIS2015-67963-P), the Grant Agency of the Czech Republic (Grant No. 18-04291S), and the IGA Project of the Palack{\'y} University (Grant No. IGA PrF 2018-003). 


\end{document}